\documentclass[12pt]{iopart}
\usepackage{float}
\usepackage{graphicx}
\usepackage{amsfonts}
\usepackage[UKenglish]{babel}
\usepackage{hyperref}
\newcommand{\Var}{\mbox{Var}}

\begin{document}
	
\title{Towards quantum key distribution with noisy communication sources}
\author{A Walton$^{1,3}$, A Ghesqui\`ere$^1$, D Jennings$^{1,2}$ and B Varcoe$^1$}
\date{\today}
\address{$^1$ School of Physics and Astronomy, University of Leeds, Leeds, LS2 9JT, United Kingdom}
\address{$^2$ Department of Physics, Imperial College London, London, SW7 2AZ, United Kingdom}
\address{$^3$ Email: pyaw@leeds.ac.uk}

\begin{abstract}
	A beam emitted by a displaced thermal source, incident on a beam splitter, is used as the source in a central broadcast quantum key distribution protocol. These displaced thermal states have parallels to signals produced through phase-shift keying, already commonly used in a variety of microwave-based communications equipment. Here, we analyse a thermal state quantum key distribution protocol for which there is no dependency on displacement. Our analysis suggests the ability to carry out QKD using widely available broadcasting equipment. We also consider the effects of the introduction of loss at several points in the protocol, and find that, while increasing loss in Alice's channel eventually results in advantage distillation being required, security is still maintained.
\end{abstract}

\maketitle

\section{Introduction}

Quantum Key Distribution (QKD) is the process by which multiple parties, here two, referred to as Alice and Bob, attempt to establish secure communication over insecure channels through applications of the principles of quantum mechanics \cite{Experimental_QKD}. For symmetric-key cryptography, this is accomplished by having Alice and Bob share a cryptographic key which can be used to encrypt and decrypt messages. Meanwhile Eve, who is eavesdropping on the channel, does not know the key. Here, the difficulty lies in initially sharing a secret key between Alice and Bob while Eve is eavesdropping.

A large section of work in QKD explores the use of coherent sources \cite{QuantumInformation} \cite{CVQKD} to produce a key, such as in the Gaussian Modulated Coherent State protocol \cite{Passive}. An advantage to this being that coherent states are accessible through lasers. However, here we will focus on the use of thermal states \cite{Thermal_1,Thermal_2,Thermal_3} which are commonly used in modern wireless communication.

Previous work involving thermal states in a central broadcast QKD protocol \cite{Thermal_QKD} showed the ability to generate a key using a beam splitter to direct a signal from an unmodulated thermal source to Alice and Bob. While interesting, communications equipment do not typically broadcast signals without some form of modulation, limiting the applicability of the previous theory in current devices.

\begin{figure}[H]
	\centering
	\includegraphics[width=0.4\textwidth]{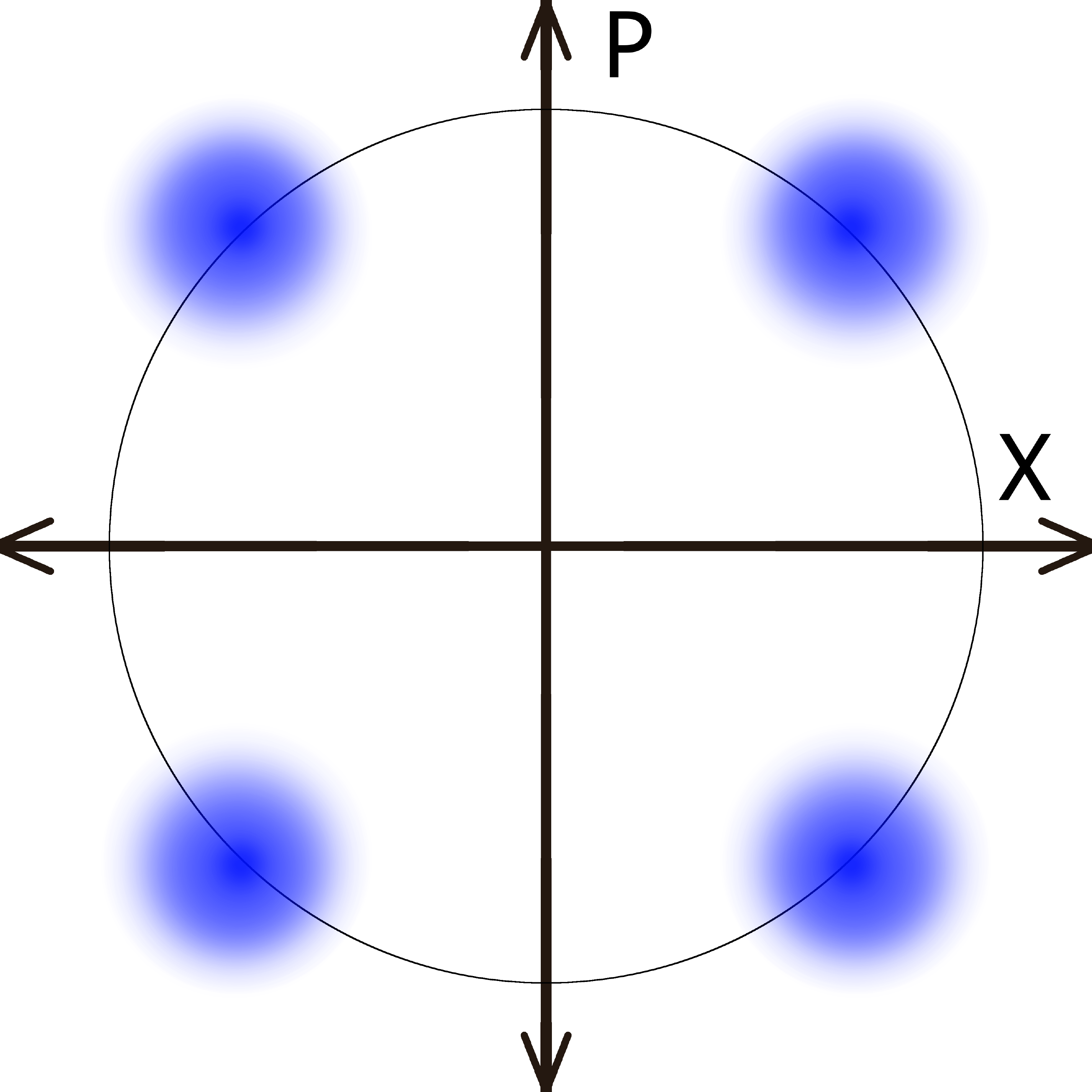}
	\caption{\textbf{Quadrature Phase-shift Keying.} A constellation diagram showing an example of PSK which sees common use in communication. Displaced thermal states are used to transmit data two bits at a time by assigning each of the four possible combinations of bit pairs to one of the clusters.
		\label{fig:QPSK}}
\end{figure}

Signal modulation is ubiquitous in communications equipment. Specifically we call attention to Phase-Shift Keying (PSK), transmitting data through modulation of the phase of a signal to one of a finite number of values, different variations of which see widespread use in areas such as Wifi, Bluetooth, satellite and mobile communication. Compared to the previous paper \cite{Thermal_QKD}, which concerned a single thermal state centred at the origin in phase space, this is instead equivalent to a series of displaced thermal states equally spaced around a circle centred at the origin. An example of one form of PSK, quadrature phase-shift keying, is shown in Figure \ref{fig:QPSK}.

Given that current microwave-based devices already broadcast displaced thermal states, it is then of interest to test if treatment of the thermal state protocol can be extended to allow for the use of displaced thermal sources, as success would suggest that QKD can be performed using widely available wireless equipment.

In this paper, Section \ref{sec:Thermal-states-and-Displacement} introduces thermal states and the displacement operator, then the protocol is described in Section \ref{sec:Protocol}. This is followed by descriptions of the mathematical modelling and results in Section \ref{sec:Modelling}. Finally, the effects of the addition of loss at various points on the key rate is analysed in Section \ref{sec:Loss}.

\section{Thermal states and displacement} \label{sec:Thermal-states-and-Displacement}

\begin{figure}[H]
	\centering
	\includegraphics[width=0.4\textwidth]{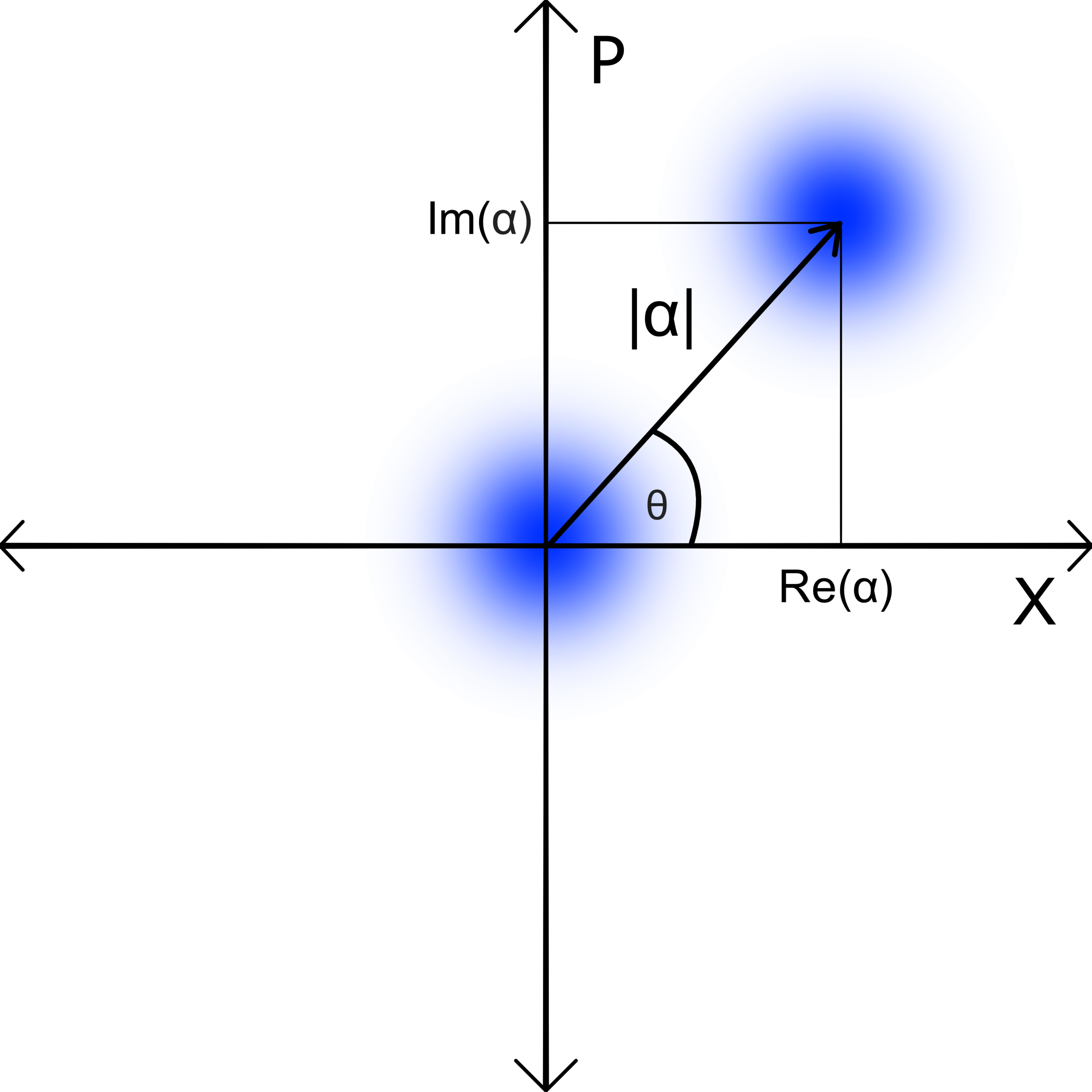}
	\caption{\textbf{Displacing a state.} Visualisation of displacing a thermal state in phase space, represented by the transformation $
		\rho_{DTh}\left(\alpha\right)=\hat{D}\left(\alpha\right)\rho_{Th}\hat{D}^{\dagger}\left(\alpha\right)$, with $\alpha=|\alpha|e^{i\theta}\in\mathbb{C} $. The distribution is translated by $\left( |\alpha|\cos\left(\theta\right),|\alpha|\sin\left(\theta\right) \right)^T$.
		\label{fig:Displacement}}
\end{figure}

If the creation and annihilation operators are given by $\hat{a}^{\dagger}$ and $\hat{a}$ respectively, the displacement operator is defined as $
\hat{D}\left(\alpha\right)=\exp\left(\alpha\hat{a}^{\dagger}-\alpha^{*}\hat{a}\right),
$ where $\alpha\in\mathbb{C} $ is the displacement parameter. The displacement operator can be applied to $\hat{a}^{\dagger}$, $\hat{a}$ \cite{Displaced_Number_States} and a thermal state $\rho_{Th}$ as shown in equations \ref{eq:DisCreation}-\ref{eq:Displaced_Thermal_State}, with equation \ref{eq:Displaced_Thermal_State} describing a displaced thermal state, shown in Figure \ref{fig:Displacement}. In phase space, the displacement operator has the effect of translating a distribution by $\left( |\alpha|\cos\left(\theta\right),|\alpha|\sin\left(\theta\right) \right)^T$, where $\alpha=|\alpha|e^{i\theta}$.

\begin{equation}
\hat{D}\left(\alpha\right)\hat{a}^{\dagger}\hat{D}^{\dagger}\left(\alpha\right)=\hat{a}^{\dagger}-\alpha^{*}\label{eq:DisCreation}
\end{equation}

\begin{equation}
\rho_{DTh}\left(\alpha\right)=\hat{D}\left(\alpha\right)\rho_{Th}\hat{D}^{\dagger}\left(\alpha\right).\label{eq:Displaced_Thermal_State}
\end{equation}

Here, the thermal state distribution is given in the Fock basis as $\rho_{Th}=\sum_{n=0}^{\infty}p_{n}|n\rangle\langle n|$, with $p_{n}=\frac{\exp\left(-n\beta\hbar\omega\right)}{1-\exp\left(-\beta\hbar\omega\right)}$, where $\beta$ is the thermodynamic beta. Unlike a coherent source \cite{Thermal_1}, it is known that a beam from a thermal source, when incident on a beamsplitter, produces a pair of output beams which have correlated intensity measurements when the output modes are compared. \cite{HBT, Photon_Bunching}. This is due to the bunched nature of photons in thermal light causing high variance in measurements of intensity. We aim to use these correlations to produce a secure key. This will be done by using a series of beamsplitters to direct a beam from a displaced thermal source to Alice, Bob and Eve. They will each perform quadrature measurements on a received beam, which will yield correlated results due to the Hanbury Brown and Twiss effect \cite{HBT}.

The protocol we will analyse is similar to others which use modulated coherent states \cite{GMCS1,GMCS2,GMCS3,GMCS4}. In these Gaussian Modulated Coherent State (GMCS) protocols, pairs of values, $x_i, y_i$ are repeatedly drawn at random from a Gaussian distribution and used by Alice to produce displaced coherent states with displacement vector $\left(x_i, y_i\right)$. These are sent to Bob for measurement, with Eve able to access the channel between Alice and Bob for interception.

From Bob and Eve's perspective, these random coherent states are indistinguishable from the output of a thermal source. Therefore, an equivalent protocol involves instead splitting the output of a thermal source, with outputs directed to Alice and Bob. The use of an actual thermal source in this protocol allows for implementation on devices which already utilise such sources.

\section{Protocol} \label{sec:Protocol}

\begin{figure}[H]
	\centering
	\includegraphics{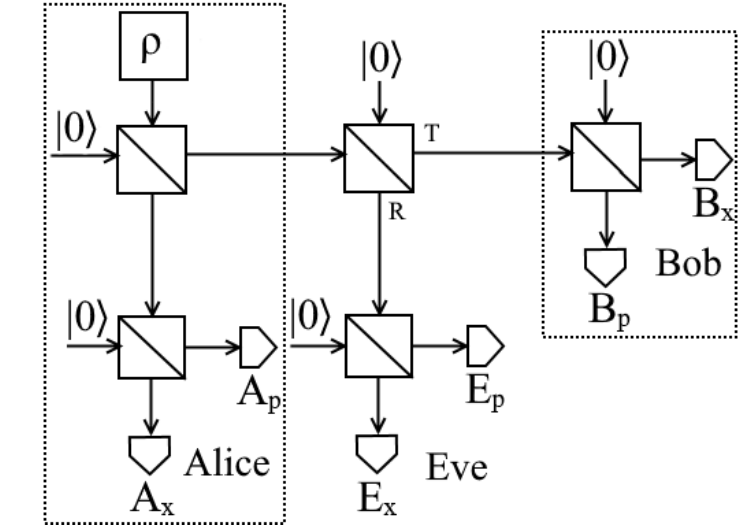}
	\caption{\textbf{The thermal state protocol.} A beam produced by a thermal source is incident on a 50:50 beamsplitter. One output is sent to Alice, with the other being sent to Bob. The beam sent towards Bob is not under Alice's control, and this insecure channel is the point in which Eve will perform the interception. This diagram was originally used in "Thermal state quantum key distribution" \cite{Thermal_QKD}, and is licensed under CC-BY 4.0 \cite{Copyright} \label{fig:Protocol}}
\end{figure}

Building on prior work \cite{Thermal_1,Thermal_QKD}, we use a central broadcast protocol in which a beam emitted from a thermal source is incident on a 50:50 beamsplitter, the outputs of which are sent to the two parties, Alice and Bob. The eavesdropper, Eve will perform the interception using an entangling cloner attack, inserting a beamsplitter of transmittance T in the channel between the initial beamsplitter and Bob. Previously \cite{Thermal_QKD}, a reduction in conditional mutual information, and therefore the key rate was observed as the reflectance of Eve's intercepting beam splitter increased. However, the produced bit strings were consistently suitable for key distillation provided the transmittance did not drop to 0\% in a noiseless scenario.

We assume that Alice is in control of the source and their own measurement apparatus, along with the channels in between them, as shown in Figure \ref{fig:Protocol}. Bob is also in control of their own measurement apparatus, but the channel leading there from the initial beamsplitter is insecure. Each person will split their received beam into a pair of outputs with a 50:50 beamsplitter, then they perform a series of pairs of measurements, $\left\{ x_{i},p_{i}\right\}$ on the X quadrature of one beam, and the P quadrature of their second beam. Here, the X and P quadratures are defined as:

\begin{equation}
\hat{X}\left(\theta\right)=\frac{1}{2}\left(\hat{a}^{\dagger}e^{i\theta}+\hat{a}e^{i\theta}\right),
\end{equation}

\begin{equation}
\hat{P}\left(\theta\right)=\frac{i}{2}\left(\hat{a}^{\dagger}e^{i\theta}-\hat{a}e^{i\theta}\right),
\end{equation}

where these quadratures are measured through heterodyne detection, by mixing the beams to be measured with a beam from a coherent source at a beamsplitter and measuring the output intensities. This is done as an alternative method to measuring in random bases, and has been successfully employed in other QKD protocols \cite{GMCS4}.
 
For each pair of measurements, each person calculates $z_{i}=\sqrt{x_{i}^{2}+p_{i}^{2}}$ to find a string of $z_{i}$ values. The median of these $z_{i}$ values are found for each person, and a value of 0 or 1 is assigned to each measurement depending on if the $z_{i}$ value is above or below that person's median. This produces a bit string for each person which may be used for key distribution.

Unlike in the previous paper \cite{Thermal_QKD}, the thermal source will be displaced to set positions in order to measure the effects of displacement on the key rate bounds. These displaced states would appear in practical implementations of the protocol, in the same way that modulated signals are used in modern microwave communication.

We do not employ random displacement \cite{CVQKDThermal} in a prepare-and-measure scheme, instead using a single thermal source displaced by a constant amount. As amplitude measurements are used to produce bit strings in this protocol, results from this single state is easily expanded to multiple thermal states separated by phase differences, in the manner shown in Figure \ref{fig:QPSK}.

The results of QKD with random displacement of thermal states in a prepare-and-measure protocol \cite{CVQKDThermal} showed a small region of secure key generation in the microwave region when transmittance to Bob was very high ($>98\%$). By replacing this with a central broadcast setup, we allow for much lower transmittance, here we perform the protocol with a transmittance of 50\% to Bob.

Previous simulations of this protocol with thermal states centred at the origin in phase space showed that, in the absence of noise and loss, the lower bound on the key rate was greater than zero provided the beamsplitter used for Eve's interception had nonzero transmittance \cite{Thermal_QKD}. This was supported by covariance matrix calculations employed to calculate von Neumann mutual information. Here, we will perform similar calculations for a state with nonzero displacement.

\section{System modelling} \label{sec:Modelling}

As mentioned in Section \ref{sec:Thermal-states-and-Displacement}, from Bob and Eve's perspective the outputs of a thermal source are equivalent to those from the sources in Gaussian Modulated Coherent State protocols. This allows the application of existing security proofs for GMCS QKD to the thermal state protocol. These include tests of composable security in the case of collective attacks \cite{Compose1, Compose2, Compose3} and finite key effects \cite{Compose3, Finite1}.

\subsection{Covariance}

The covariance matrix of the system can be used to calculate entropy, and therefore the bounds on the key rate. In this protocol, we can completely describe the state of the system through the first and second moments as there are only Gaussian modes involved. For an N-mode system, the covariance matrix is defined as \cite{Raul}:

\begin{equation}
\gamma_{ij}=\Tr\left[\rho\frac{1}{2}\left\{ \left(\hat{r}_{i}-d_{i}\right),\:\left(\hat{r}_{j}-d_{j}\right)\right\} \right],
\end{equation}

\begin{equation}
d_{i}=\left\langle \hat{r}_{i}\right\rangle ,
\end{equation}

\begin{equation}
r=\left(\hat{X}_{1},\:\hat{P}_{1},\:...,\:\hat{X}_{N},\:\hat{P}_{N}\right).
\end{equation}

Here, $\hat{X}_{i},$ and$\:\hat{P}_{i}$ are the pair of quadrature operators for mode $i$ of the system and $\{a,b\}=ab+ba$ is the anticommutator. When considering only the two modes incident on the initial beamsplitter, the input for a displaced thermal source is given by:

\begin{equation}
d_{1}=\left[\begin{array}{c}
d_{x}\\
d_{p}\\
v_{1x}\\
v_{1p}
\end{array}\right],
\end{equation}

where $d_{x}$ and $d_{p}$ describe the quadratures of the source beam. Vacuum noise added in the second input is described by $v_{1x}$ and $v_{1p}$. The beamsplitters shown in the protocol are applied in succession to the relevant modes through the use of the beamsplitter transformation $S\left(T,R\right)$:

\begin{equation}
S\left(T,R\right)=\left[\begin{array}{cc}
T & R\\
-R & T
\end{array}\right]\otimes I.
\end{equation}

Here, $T$ and $R$ are the transmittance and reflectance of the beamsplitter, defined such that $T^{2}+R^{2}=1$. With the exception of Eve's beamsplitter, which has unknown transmittance, each of the beamsplitters used in this protocol will be 50:50, with $T^{2}=R^{2}=\frac{1}{2}.$ We can find the final output vector and covariance matrix of the system. Considering only the X-quadratures, we can find half of the output vector:

\begin{equation}
r_{x}=\left[\begin{array}{c}
\frac{t_{a}}{2}\left(d_{x}+v_{1x}\right)+\frac{t_{a}}{\sqrt{2}}v_{2x}+r_{a}v_{A_{1}}\\
-\frac{t_{a}}{2}\left(d_{x}+v_{1x}\right)+\frac{t_{a}}{\sqrt{2}}v_{2x}+r_{a}v_{A_{2}}\\
-\frac{Tt_{b}}{2}\left(d_{x}-v_{1x}\right)+\frac{Rt_{b}}{\sqrt{2}}v_{3x}+\frac{t_{b}}{\sqrt{2}}v_{4x}+r_{b}v_{B_{1}}\\
\frac{Tt_{b}}{2}\left(d_{x}-v_{1x}\right)-\frac{t_{b}R}{\sqrt{2}}v_{3x}+\frac{t_{b}}{\sqrt{2}}v_{4x}+r_{b}v_{B_{2}}\\
\frac{t_{e}R}{2}\left(d_{x}-v_{1x}\right)+\frac{Tt_{e}}{\sqrt{2}}v_{3x}+\frac{t_{e}}{\sqrt{2}}v_{5x}+r_{e}v_{E_{1}}\\
-\frac{t_{e}R}{2}\left(d_{x}-v_{1x}\right)-\frac{Tt_{e}}{\sqrt{2}}v_{3x}+\frac{t_{e}}{\sqrt{2}}v_{5x}+r_{e}v_{E_{2}}
\end{array}\right].
\end{equation}

\begin{figure}[H]
	\centering
	\includegraphics{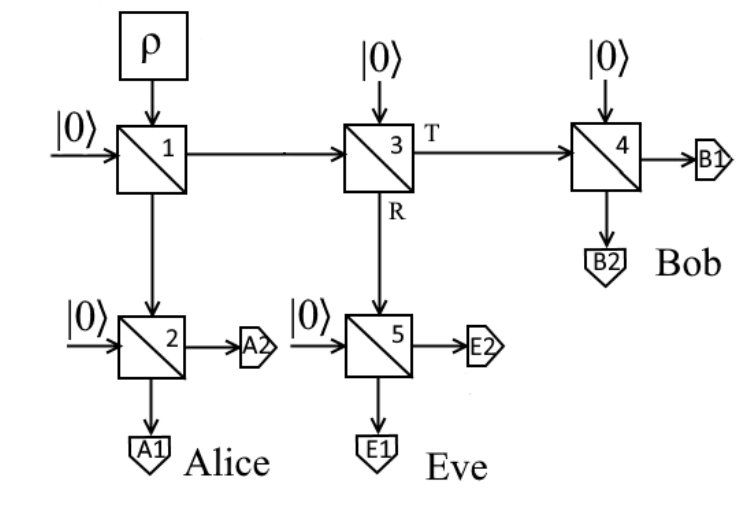}
	\caption{\textbf{Beamsplitter and detector labels.}  The noise added at each beamsplitter is described by $v_{1-5}$ while the detector noises are given by $v_{A_{1}},\:v_{A_{2}},\:v_{B_{1}},\:v_{B_{2}},\:v_{E_{1}},\:v_{E_{2}}$. \label{fig:BeamsplitterLabels}}
\end{figure}

The P-quadrature results, $r_{p}$, follow the same pattern, giving a final output vector of $r_{2}=\left[r_{x1},r_{p1},r_{x2},r_{p2},...,r_{x6},r_{p6}\right]^{T}$. Eve performs the interception using a beamsplitter with transmittance $T$ and reflectance $R$. Here, $v_{1-5}$ describe vacuum noise added by the beamsplitters in the order labelled in Figure \ref{fig:BeamsplitterLabels}.

Detector efficiency is modelled using an additional beamsplitter at each of the six detectors, with added vacuum noises $v_{A_{1}},\:v_{A_{2}},\:v_{B_{1}},\:v_{B_{2}},\:v_{E_{1}},$ and $v_{E_{2}}$. The beamsplitters modelling Alice's pair of detectors are both assumed to have transmittance $t_{a}$ and reflectance $r_{a}$, allowing loss to be introduced during detection. To model detection loss at Bob and Eve's pairs of detectors, $t_{b},\:r_{b},\:t_{e},$ and $r_{e}$ are similarly defined.

We can calculate the covariance matrix using the final vector. We set $v=\Var\left(d_{x}\right)=\Var\left(d_{p}\right)$ to be the variance of the initial state, and take the added noise to have mean zero and variance one. This yields:

\begin{equation}
\gamma_{A_{1}A_{2}B_{1}B_{2}E_{1}E_{2}}=\left[\begin{array}{ccc}
\gamma_{A_{1}A_{2}} & C_{AB} & C_{AE}\\
C_{AB}^{T} & \gamma_{B_{1}B_{2}} & C_{BE}\\
C_{AE}^{T} & C_{BE}^{T} & \gamma_{E_{1}E_{2}}
\end{array}\right]
,\label{eq:CovMatrix}
\end{equation}

Where the sub-matrices are given by:

\begin{equation}
\gamma_{A_{1}A_{2}}=\left[\begin{array}{cc}
\frac{t_{a}^{2}}{4}\left(v+3\right)+r_{a}^{2} & \frac{t_{a}^{2}}{4}\left(1-v\right)\\
\frac{t_{a}^{2}}{4}\left(1-v\right) & \frac{t_{a}^{2}}{4}\left(v+3\right)+r_{a}^{2}
\end{array}\right]\otimes I
\end{equation}

\begin{equation}
\gamma_{B_{1}B_{2}}=\left[\begin{array}{cc}
\frac{T^{2}t_{b}^{2}}{4}\left(v+1\right)+\frac{t_{b}^{2}\left(r^{2}+1\right)}{2}+r_{b}^{2} & -\frac{T^{2}t_{b}^{2}}{4}\left(v+1\right)+\frac{t_{b}^{2}\left(1-R^{2}\right)}{2}\\
-\frac{T^{2}t_{b}^{2}}{4}\left(v+1\right)+\frac{t_{b}^{2}\left(1-R^{2}\right)}{2} & \frac{T^{2}t_{b}^{2}}{4}\left(v+1\right)+\frac{t_{b}^{2}\left(R^{2}+1\right)}{2}+r_{b}^{2}
\end{array}\right]\otimes I
\end{equation}

\begin{equation}
\gamma_{E_{1}E_{2}}=\left[\begin{array}{cc}
\frac{t_{e}^{2}R^{2}}{4}\left(v+1\right)+\frac{t_{e}^{2}\left(T^{2}+1\right)}{2}+r_{e}^{2} & -\frac{R^{2}t_{e}^{2}}{4}\left(v+1\right)+\frac{t_{e}^{2}\left(1-T^{2}\right)}{2}\\
-\frac{R^{2}t_{e}^{2}}{4}\left(v+1\right)+\frac{t_{e}^{2}\left(1-T^{2}\right)}{2} & \frac{t_{e}^{2}R^{2}}{4}\left(v+1\right)+\frac{t_{e}^{2}\left(T^{2}+1\right)}{2}+r_{e}^{2}
\end{array}\right]\otimes I
\end{equation}

\begin{equation}
C_{AB}=\left[\begin{array}{cc}
\frac{Tt_{a}t_{b}}{4}\left(1-v\right) & \frac{Tt_{a}t_{b}}{4}\left(v-1\right)\\
\frac{Tt_{a}t_{b}}{4}\left(v-1\right) & \frac{Tt_{a}t_{b}}{4}\left(1-v\right)
\end{array}\right]
\end{equation}

\begin{equation}
C_{AE}=\left[\begin{array}{cc}
\frac{Rt_{a}t_{e}}{4}\left(v-1\right) & \frac{Rt_{a}t_{e}}{4}\left(1-v\right)\\
\frac{Rt_{a}t_{e}}{4}\left(1-v\right) & \frac{Rt_{a}t_{e}}{4}\left(v-1\right)
\end{array}\right]
\end{equation}

\begin{equation}
C_{BE}=\left[\begin{array}{cc}
\frac{TRt_{b}t_{e}}{4}\left(1-v\right) & \frac{TRt_{b}t_{e}}{4}\left(v-1\right)\\
\frac{TRt_{b}t_{e}}{4}\left(v-1\right) & \frac{TRt_{b}t_{e}}{4}\left(1-v\right)
\end{array}\right]
\end{equation}

This describes the covariances of the modes received by Alice, Bob, then Eve's pairs of detectors.

\subsection{Final State Entropy}

Using the covariance matrix, we can calculate von Neumann joint entropy between pairs of modes, which can be used to place bounds on the key rate in the case of a collective attack by Eve. Here, Eve stores their received states in memory, and performs measurements after all classical communication concerning privacy amplification between Alice and Bob has ended. The von Neumann entropy of a state  $\rho$ is given by \cite{Raul}:

\begin{equation}
S\left(\rho\right)=\sum_{i}G\left(\frac{\lambda_{i}-1}{2}\right),\label{eq:SSummation}
\end{equation}

\begin{equation}
G\left(x\right)=\left(x+1\right)\log_{2}\left(x+1\right)-x\log_{2}x,
\end{equation}

where $\lambda_{i}$ denotes the symplectic eigenvalues of the final state of the system.
 Here, $\lambda^{2}=\left|\gamma_{1}\right|$ for a one mode system, while for two modes we can calculate: 

\begin{equation}
\left(\lambda_{1,2}\right)^{2}=\frac{1}{2}\left(\Delta\pm\left[\Delta^{2}-4\left|\gamma_{12}\right|\right]^{\frac{1}{2}}\right),
\end{equation}

\begin{equation}
\Delta=\left|\gamma_{1}\right|+\left|\gamma_{2}\right|-2\left|C\right|.\label{eq:Delta}
\end{equation}

By defining the mutual information between two modes as $I\left(A;B\right)=S\left(A\right)+S\left(B\right)-S\left(AB\right),$ and the conditional mutual information as $I\left(A;B|E\right)=S\left(AE\right)+S\left(BE\right)-S\left(ABE\right)-S\left(E\right)$ we can place limits on the key rate, $K\left(A;B|E\right)$ \cite{Key_Rate}.

\begin{equation}
K\left(A;B|E\right)\geq\max\left[I\left(A;B\right)-I\left(A;E\right),\,I\left(A;B\right)-I\left(B;E\right)\right],
\end{equation}

\begin{equation}
K(A;B|E)\leq\min\left[I\left(A;B\right),\,I\left(A;B|E\right)\right],
\end{equation}

From equations \ref{eq:SSummation}-\ref{eq:Delta}, as well as the covariance matrix shown in equation \ref{eq:CovMatrix}, we find that displacement does not have any effect on the key rate bounds. Dependency on $d_{x}$ and  $d_{p}$ disappears in the final matrix, leaving only dependency on the thermal state variance. Therefore, calculating mutual information values using elements of the covariance matrix will produce limits which are also not dependent on displacement. From this we can see that the protocol can be carried out without requiring a source with zero displacement. Having shown that the protocol functions in a no-loss scenario with displaced thermal states, we can consider cases where loss is present.

\section{Loss} \label{sec:Loss}

\begin{figure}[H]
	\centering
	\includegraphics[width=0.8\textwidth]{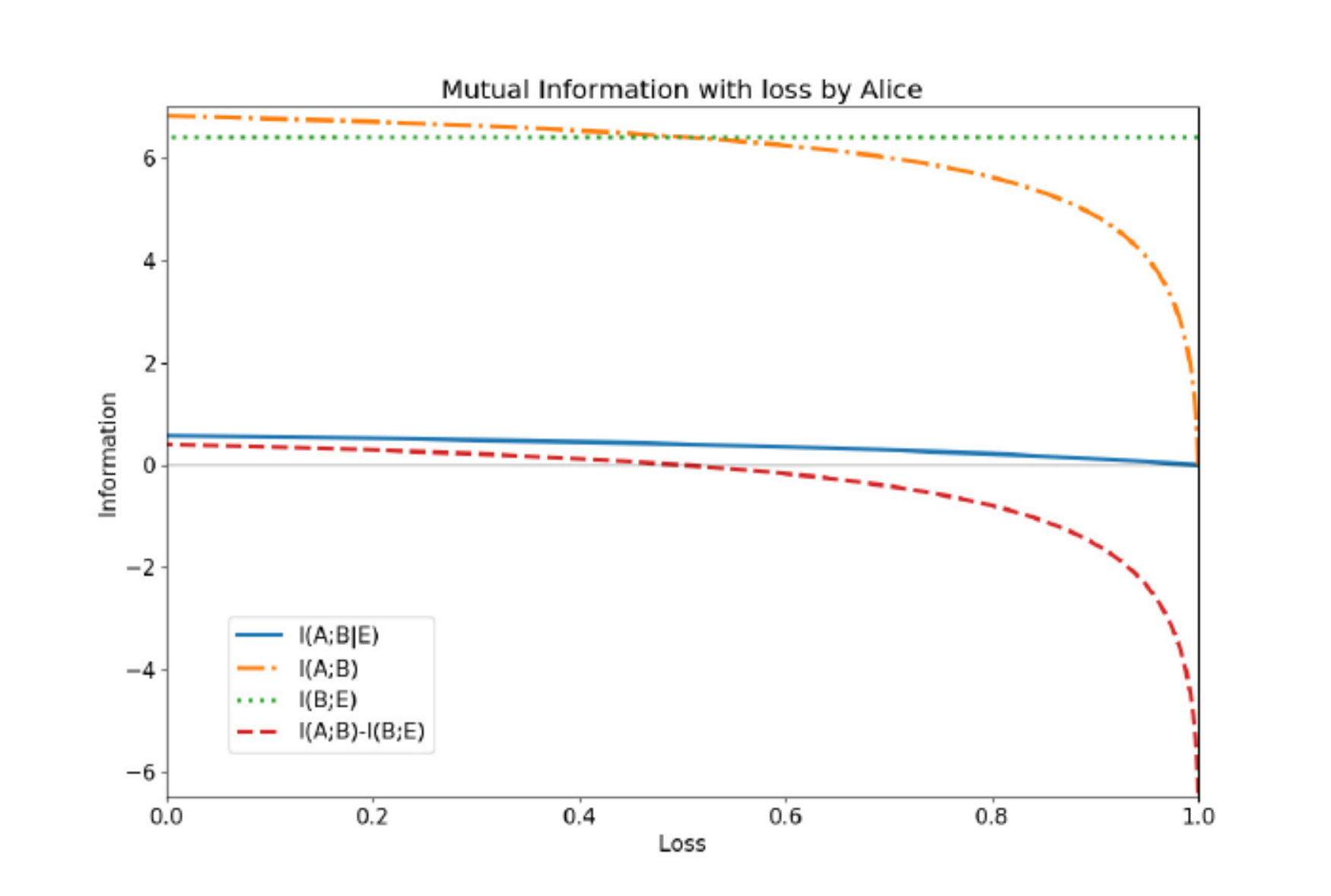}
	\caption{\textbf{Alice's loss.} Changes in von Neumann mutual information as loss is introduced at Alice's detectors, calculated using the covariance matrix. $K=I\left(A;B\right)-I\left(B;E\right)$ becomes negative before total loss is reached. \label{fig:AliceLoss}}
\end{figure}

\begin{figure}[H]
	\centering
	\includegraphics[width=0.8\textwidth]{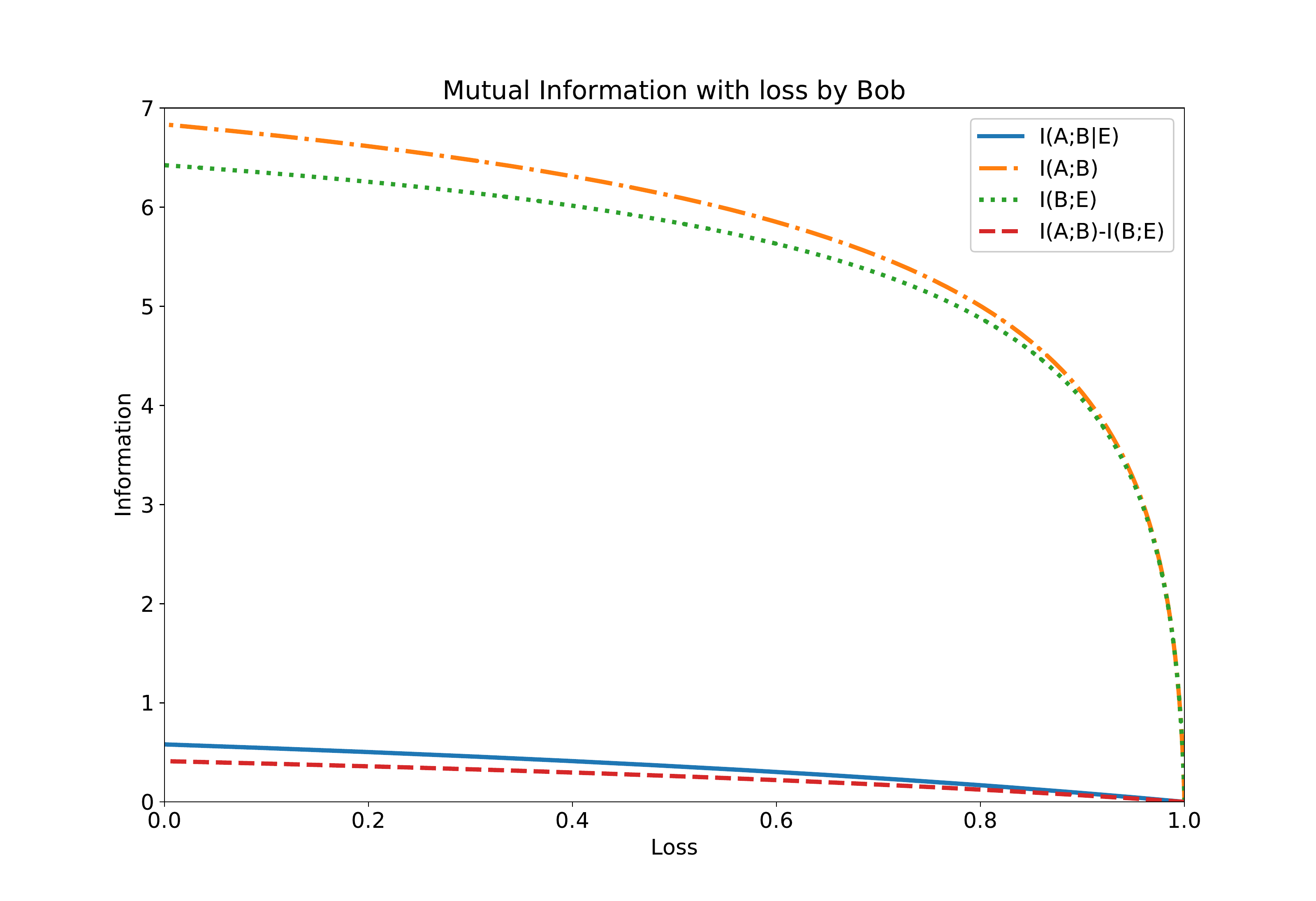}
	\caption{\textbf{Bob's loss.} Changes in von Neumann mutual information as loss is introduced at Bob's detectors, calculated using the covariance matrix. As long as Bob receives some of the signal sent to him, a key can be produced. \label{fig:BobLoss}}
\end{figure}

\begin{figure}[H]
	\centering
	\includegraphics[width=0.8\textwidth]{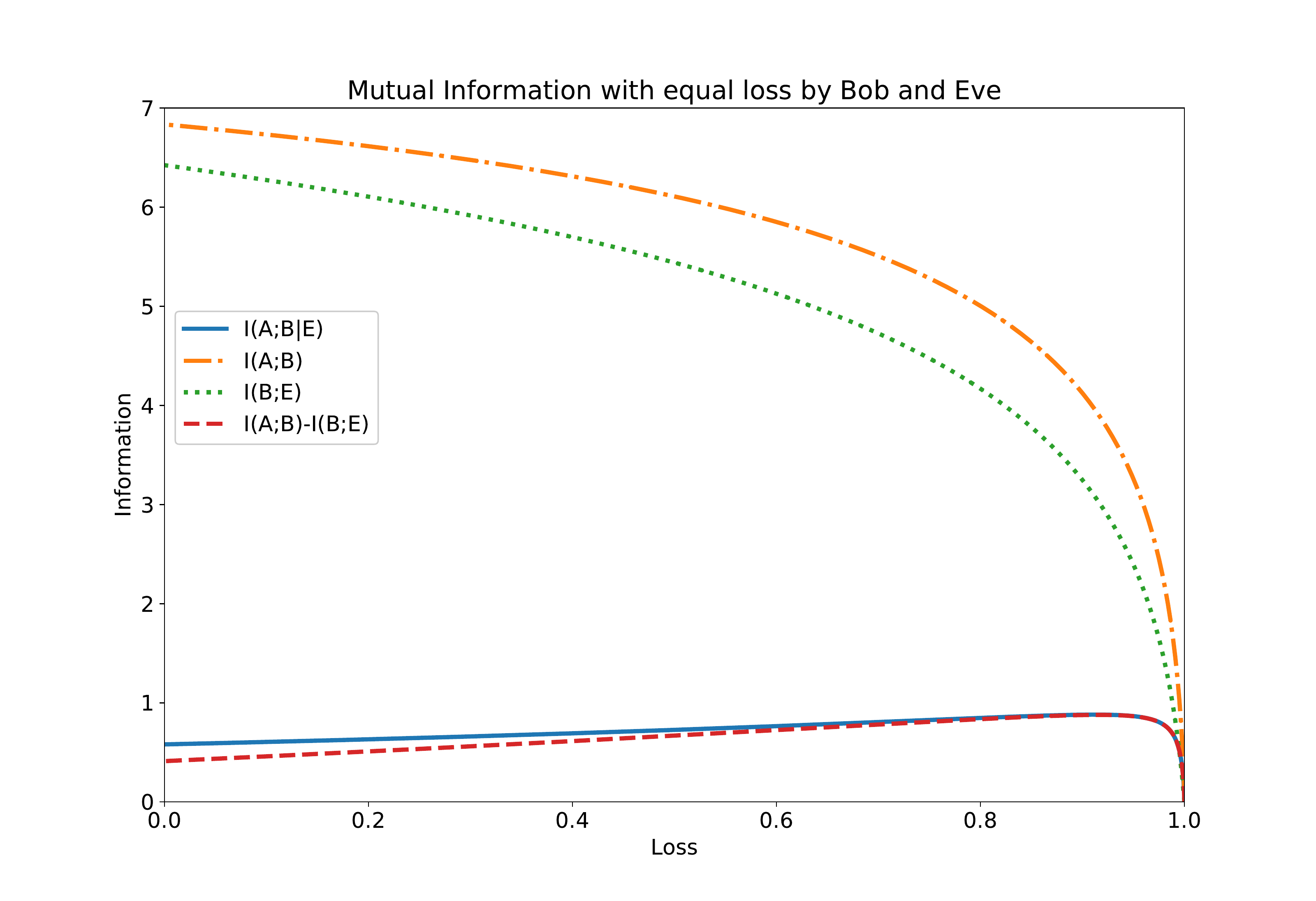}
	\caption{\textbf{Loss before interception.} Changes in von Neumann mutual information with equal loss added at Bob and Eve's detectors, this simulates loss added before Eve's interception. In this final case, communication is secure for nonzero transmittance. This is relevant for wireless scenarios, as then Eve will not realistically be able to detect every part of the signal that is not detected by Alice and Bob. \label{fig:BobEveLoss}}
\end{figure}

We now analyse the effects of loss in the system, with Eve performing interception using a 50:50 beamsplitter. Here, we choose to model loss through the use of beam splitters at the detectors. This allows us to apply loss to Alice's beam before measurement can occur, as well as on Bob's beam before or after Eve performs interception. We consider the value $K=I\left(A;B\right)-I\left(B;E\right)$, a lower bound on the key rate, due to $K$ being a positive function when testing without loss, with $K>0$ being a requirement \cite{Bounds} for QKD to take place via reverse reconciliation. This is calculated using equation \ref{eq:SSummation}. Also considered in the conditional mutual information, as secrecy can be ensured through advantage distillation provided conditional mutual information is positive \cite{Maurer99}. The results of these calculations for the three scenarios are displayed in Figures \ref{fig:AliceLoss}-\ref{fig:BobEveLoss}. When loss is introduced into Alice's channel, Alice and Bob will eventually no longer have the superior channel, with $K=0$ occurring at approximately 50\% loss. However when loss is introduced into Bob's channel, either before or after Eve performs interception, $K=I\left(A;B\right)-I\left(B;E\right)$ remains positive.

In the case where Eve's channel is superior, advantage distillation \cite{Key_Rate}\cite{Advantage1} can be used. There, an authenticated public channel is used by Alice and Bob to improve their mutual information. Otherwise, reverse reconciliation is sufficient to produce a key.

\begin{figure}[H]
	\centering
	\includegraphics[width=0.8\textwidth]{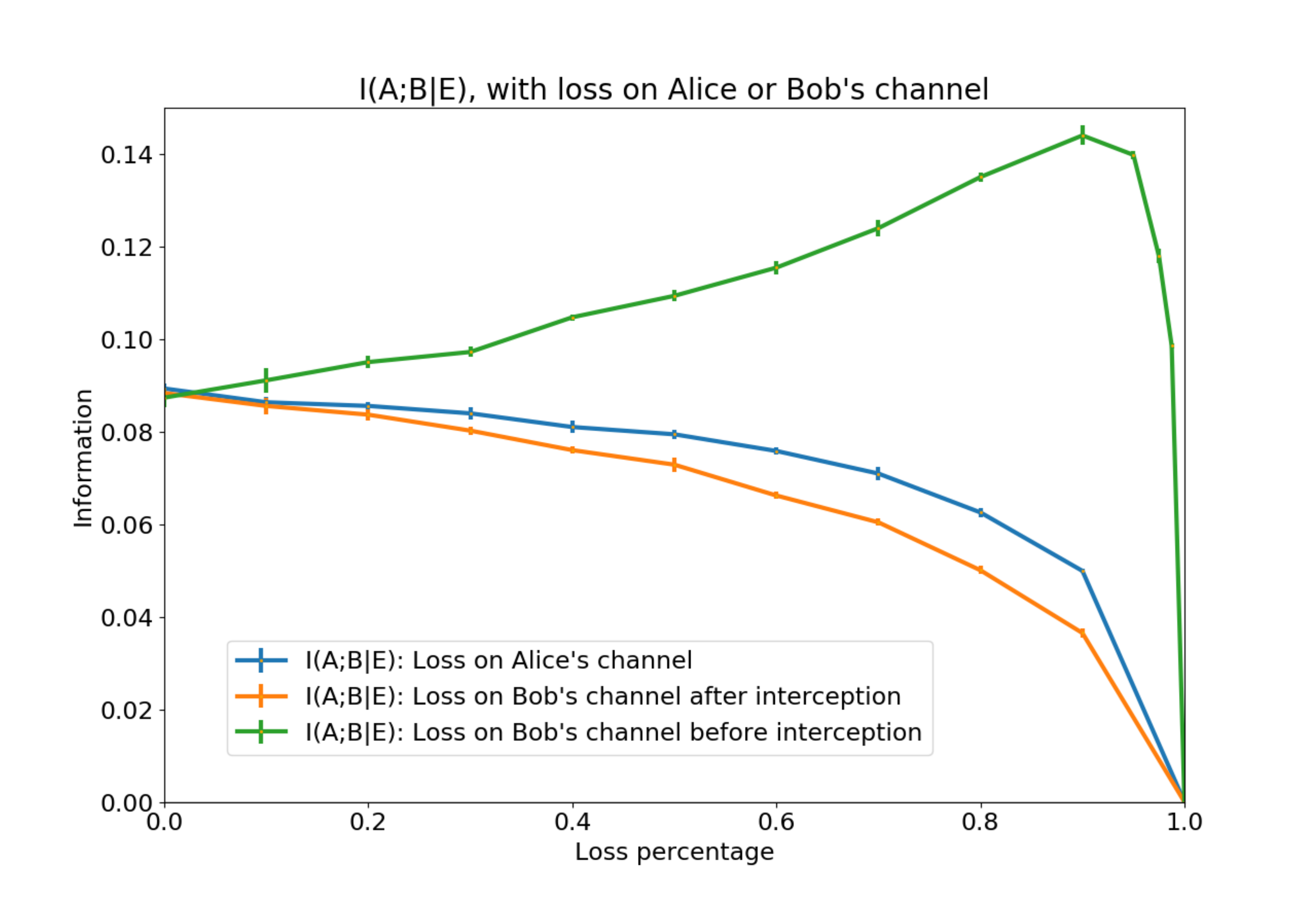}
	\caption{\textbf{Simulation results.} Shannon mutual information measurements \cite{Dataset} calculated using bit strings produced through simulation, these each follow the trends predicted by the covariance matrix calculations. \label{fig:Mutual-information-Simulation}}
\end{figure}

These results were compared to the outputs of Monte Carlo simulations of the protocol that were performed using QuTiP \cite{QuTiP1,QuTiP2}, with Shannon entropy calculations performed on bit strings derived from measurements recorded by Alice, Bob and Eve. When loss is introduced on Alice's channel, or in Bob's channel before or after interception, as shown in Figure \ref{fig:Mutual-information-Simulation}, changes in conditional mutual information values derived from bit strings mirror the curves calculated through the covariance matrix, given in Figures \ref{fig:AliceLoss}-\ref{fig:BobEveLoss}. As in previous work \cite{Thermal_QKD}, it is expected that the von Neumann entropy is of greater magnitude, due to the presence of discord in the system.

\section{Conclusions} \label{sec:Conclusion}

Calculating the final covariance matrix of the system shows that the dependency on displacement, observed in the vector describing the input state, is removed. This means that the bounds on the key rate also have no displacement dependency. Therefore the thermal state protocol can be generalised to include displaced thermal sources without sacrificing the key rate or security. When loss was introduced into the system for Alice, Alice and Bob were eventually put into an inferior position compared to Eve, so effort should be made to minimise loss across this section of the system in experimental work. However, loss in any case still allowed a key to be produced through advantage distillation. In future work, the protocol will be implemented experimentally using software-defined radios, allowing realistic noise and loss to be described. This would allow the creation of a real key via privacy amplification, showing it is experimentally possible to produce keys using this protocol using easily accessible equipment.																														

\section*{Acknowledgements}

This work was undertaken on ARC4, part of the High Performance Computing facilities at the University of Leeds, UK. DJ is supported by the Royal Society and also a University Academic Fellowship. The data used in Figure \ref{fig:Mutual-information-Simulation} is available at \url{https://doi.org/10.5518/1056} \cite{Dataset}.

\appendix

\section*{References}

\bibliographystyle{unsrt}
\bibliography{References}

\end{document}